\documentclass[12pt]{article}
\usepackage{latexsym}
\usepackage{graphicx}
\begin{document}
\title{The Semiclassical Description of Tunneling in Scattering
with Multiple Degrees of Freedom.}
\author{G.F. Bonini$^a$, A.G. Cohen$^b$, C. Rebbi$^b$ and
  V.A. Rubakov$^c$\ \thanks{\tt bonini@thphys.uni-heidelberg.de, 
    cohen@bu.edu, rebbi@bu.edu, rubakov@ms2.inr.ac.ru}\\ \\
  \small \sl   $^a$Institut f\"ur theoretische Physik, 
University of Heidelberg,\\ 
  \small \sl D-69120 Heidelberg, Germany\\
  \small \sl   $^b$Department of Physics, Boston University,
  Boston, MA 02215, USA \\
  \small \sl   $^c$Institute for Nuclear Research of the Russian Academy
of Sciences,\\
  \small \sl Moscow, 117312, Russian Federation  \\
  }

\date{}
\maketitle

\begin{abstract}
  We describe a computational investigation of 
  tunneling at finite energy in a weakly coupled 
  quantum mechanical system with two degrees of freedom. 
  We compare a full quantum mechanical analysis to the results 
  obtained by making use of a semiclassical technique developed
  in the context of instanton-like transitions in quantum field
  theory.  This latter technique is based on an analytic continuation
  of the degrees of freedom into a complex phase space, and the
  simultaneous analytic continuation of the equations of motion into
  the complex time plane.
\end{abstract}

\section{Introduction and Motivation}
  \label{sec:intro}

The existence of a small parameter (``coupling constant'') in 
quantum mechanical systems leads to a
(typically asymptotic) expansion of observables in powers of
this parameter. No other approximation technique has proved as
powerful in obtaining physical predictions in such diverse
fields as  atomic physics, chemistry, quantum field theory, {\it etc.} 
Despite these successes, many phenomena in such systems
are not amenable to perturbation theory: for example barrier penetration
in quantum mechanics does not occur  at any order in an (asymptotic)
expansion in powers of $\hbar$. 

Techniques for dealing with non-perturbative phenomena in theories
with a small parameter are far less general. Perhaps the best known
example is the WKB approximation, familiar from one-dimensional
wave mechanics. A similar technique, the
instanton method, is often used to discuss certain non-perturbative
phenomena in quantum field theory. But in more complicated cases
these methods fail: examples are tunneling at non-zero energy
in quantum mechanical systems with more than one degree of freedom, and
tunneling processes in quantum field-theoretic models with exclusive
initial or final states. 

Recently the authors of Ref.~\cite{Rubakov:1992fb,Rubakov:1992ec} have
suggested a 
method for dealing with high-energy processes that proceed
through tunneling in weakly coupled quantum field theory.  Their
technique begins with a path integral representation of the matrix
element in question, followed by a double analytic continuation: the
fields in the path integral are continued to the complex plane, and in
addition the time evolution is continued along a complex contour.
In spite of these complications, this technique is essentially
semiclassical. The resulting complexified
classical system typically remains intractable to analytic
methods. Consequently computational techniques must be employed to
obtain quantitative results; the feasibility of the corresponding
calculations in field theory has been demonstrated 
in Ref.~\cite{Kuznetsov:1997az,Kuznetsov:1996cm}. 
We should also stress that the validity of the formalism of
Ref.~\cite{Rubakov:1992fb,Rubakov:1992ec} has not been proven, though its
plausibility has been supported by comparison with perturbative
calculations about the instanton \cite{Tinyakov:1992fn,Mueller:1993sc}.

More specifically, the process under discussion is a non-perturbative
instanton mediated transition induced by the collision of two highly
energetic particles.  Perturbative calculations (see for instance
Ref.~\cite{Mattis:1992bj,Tinyakov:1993dr,Rubakov:1996vz} and
references therein) about the instanton 
suggest that the total cross section has the following functional form
\begin{equation}
  \label{eq:twotoany}
    \sigma_{2 \to {\rm any}} \propto e^{ -{1 \over g^2} F_0(g^2 E)}
\end{equation}
where $g$ is the small coupling constant of the theory and $E$ is the 
center-of-mass energy.  To compute the leading exponent
the authors of Ref.~\cite{Rubakov:1992fb,Rubakov:1992ec} 
suggested considering an inclusive process with a large
number $n$ of incoming particles. They argued that
the total probability has a similar form
\begin{equation}
  \label{eq:manytoany}
    \sigma_{n \to {\rm any}} \propto e^{ -{1 \over g^2} 
    F(g^2 E, g^2 n)}
\end{equation}
and that the exponent $F(g^2 E, g^2 n)$ can be calculated
semiclassically by considering a complexified classical system.
Furthermore, they conjectured that the two-particle exponent
$F_0$ in Eq.~(\ref{eq:twotoany}) is an appropriate limit of the 
multi-particle one:
\begin{equation}
  \label{eq:fzero}
    F_0(g^2E)={\rm lim}_{g^2n \to 0} F(g^2E,g^2n)       
\end{equation}
Equations~(\ref{eq:twotoany}),~(\ref{eq:fzero}) on the one hand, and
equation~(\ref{eq:manytoany}) on the other,
have different status. While the
validity of Eq.~(\ref{eq:manytoany}) has been demonstrated by
path integral methods~(cfr. Section 4), neither the general
functional form (\ref{eq:twotoany}) nor the limiting procedure 
(\ref{eq:fzero}) have been proven so far.

Since the formalism of Ref.~\cite{Rubakov:1992fb,Rubakov:1992ec}
has not been rigorously derived from first 
principles, and the direct evaluation of the resulting path integral,
by computer simulation or other numerical procedures is beyond
current reach, we have chosen to test the technique
by reducing the number of degrees of freedom. In quantum field 
theoretic models, a general 
field configuration may be expanded in a complete
(infinite) basis of normal modes.  In the asymptotic time domains
$t \to \pm \infty$ these modes are non-interacting, and the evolution
is characterized by a definite particle number.  In a semiclassical
description of the tunneling process the field evolves through
a non-linear regime, and the crucial question is how the
particle numbers in the incoming and outgoing asympotic states
are related by the non-linear evolution.  A minimal model
capable of mimicking this dynamics will have some internal
degree of freedom, whose excitations at asymptotic times
will correspond to the particle number of the field theoretical
system, and a non-linear interaction with a barrier,
that can be penetrated by tunneling.  This can be realized
with a system of two particles moving in one dimension.  Let
the coordinates of these particles be
$x_1$ and $x_2$, and the dynamics be described by the Lagrangian:
\begin{equation}
  \label{eq:prelagrangian}
    L = {1\over 4} {\dot x_1}^2 + {1\over 4} {\dot x_2}^2 - {1\over 8}
    \omega^2 (x_1-x_2)^2 - V(x_1)
\end{equation}
where $V$ is an arbitrary positive semi-definite potential which
vanishes asymptotically\footnote{We could 
of course allow $V$ to depend on $x_2$ as well, provided it does not
depend only on the combination $x_1-x_2$.}.
Since the 
theory is to be weakly coupled, we assume a potential of the form 
\begin{equation}
  \label{eq:potential}
    V(x) = {1 \over g^2} U(g\, x)
\end{equation}
with $g \ll 1$.
For simplicity we will use a gaussian for the potential 
\begin{equation}
  \label{eq:gaussian}
    U(x) \equiv e^{-\frac{1}{2}x^2}
\end{equation} 
although the treatment of other potentials is similar.  The properties
of the system described by the above Lagrangian are made clearer by
replacing the variables $x_1, x_2$ with the center of mass coordinate
$X \equiv (x_1+x_2)/2$ and the relative coordinate $y \equiv (x_1 -
x_2)/2$.  With this substitution the Lagrangian takes the form 
\begin{equation}
  \label{eq:lagrangian}
    L = \frac{1}{2} {\dot X}^2 + \frac{1}{2} {\dot
    y}^2 - \frac{1}{2} \omega^2 y^2 - 
    \frac{1}{g^2} e^{-\frac{1}{2}g^2 (X+y)^2}
\end{equation}
and we see that asymptotically it describes the free motion
of the center of mass and a decoupled harmonic
oscillator.  Within the range of the potential, though,
the two degrees of freedom are coupled, giving
rise to a transfer of energy between them.

In the classical case, the coupling $g$ is an irrelevant parameter:
we may rescale the degrees of freedom so that $g$ appears as a 
universal multiplicative factor. Defining new coordinates 
$\tilde X \equiv g X,\; \tilde y \equiv g y$  the Lagrangian becomes
\begin{equation}
  \label{eq:rescaledlgn}
    L = \frac{1}{g^2}\Biggl[\frac{1}{2} {\dot{\tilde X}}^2 + \frac{1}{2} 
    {\dot{\tilde y}}^2 - \frac{1}{2} \omega^2 \tilde y^2 - 
    e^{-\frac{1}{2}(\tilde X+\tilde y)^2}\Biggr] 
\end{equation}
The value of $g$ is crucial, however,
for the quantum system:  the path integral formulation
of quantum mechanics together with Eq.~(\ref{eq:rescaledlgn}) show
that $g^2$ plays a role similar to $\hbar$ in determining the
magnitude of the quantum fluctuations; the classical limit
corresponds to $g^2 \to 0$.  This is in close analogy with the
field theoretical systems mentioned above.  In the following
we will use units with $\hbar=1$ and will characterize
the semiclassical treatment as an expansion for small $g$.

The repulsive potential implies a barrier that must be either overcome
or penetrated through tunneling for a transition from an initial
state where the center of mass coordinate is approaching the barrier from, 
{\it e.g.}, large negative $X$, to a final state where it is moving 
away from it towards large positive $X$.  The corresponding transmission
probability $\cal T$ will depend on three quantities: the total initial energy
$E$; the initial energy of the oscillator $E_{osc}$ (or, equivalently,
on its initial quantum number $n$ related to its energy by
$E_{osc}=(n+1/2) \omega$); and the value of $g$.  In analogy with 
Eq.~(\ref{eq:twotoany}), the transmission probability for 
an oscillator initially in its ground state as $g \to 0$
has the asymptotic form  
\begin{equation}
  \label{eq:transzero}
    {\cal T}_0(E)=C_0(g^2 E) e^{-\frac{1}{g^2} F_0(g^2 E)}
\end{equation}
for some prefactor $C_0$.  Likewise, for a transition from an
initial state in the $n$-th excited level one expects, for $g \to
0$, $n g^2$ fixed, 
\begin{equation}
  \label{eq:transn}
    {\cal T}_n(E)=C(g^2 E, g^2 n) e^{-\frac{1}{g^2} F(g^2 E, g^2 n)}
\end{equation}
in analogy to Eq.~(\ref{eq:manytoany}).  The advantage of the
model we are considering is that the process in question admits
a full quantum mechanical treatment as well as the semiclassical
analysis.  We will present the results of a numerical solution
of the full Schr\"odinger equation and show that the transition from 
the oscillator ground state can indeed be fitted very well
with the expression of Eq.~(\ref{eq:transzero}).  Independently of
the semiclassical analysis, this result represents a direct
verification of the functional form of Eq.~(\ref{eq:twotoany}),
(\ref{eq:transzero}). This will be our first conclusion. 

We will then use the technique of Ref.~\cite{Rubakov:1992fb,Rubakov:1992ec}
and evaluate the function $F(g^2 E, g^2 n)$ entering Eq.~(\ref{eq:transn})
by solving numerically the complexified classical equations
on the appropriate contour in the complex time plane.  We will
thus be able to check the validity of Eq.~(\ref{eq:fzero}), with the 
l.h.s., $F_0(g^2 E)$, obtained through the full quantum mechanical 
treatment and the r.h.s., $F(g^2 E, g^2n)$, calculated in a semiclassical 
way.  We will show that Eq.~(\ref{eq:fzero}) indeed holds, and so, in
the context of our model at least, we will be able to confirm the 
conjecture of Ref.~\cite{Rubakov:1992fb,Rubakov:1992ec} by a direct 
numerical computation. This will be the second main conclusion of this paper.

\section{The Classical System}
  \label{sec:classical}

Let us first consider a classical evolution whereby the two
particles are initially located on the negative $x$-axis well outside
the range of the potential and their center of mass is moving with
positive velocity ({\it i.e.}~toward the barrier).  The motion of the system
is specified completely by four initial value data.  Time
translation invariance of the system allows us to choose one 
of these to be the initial time.
It is convenient to take the remaining three to be the
rescaled total energy of the system, $\epsilon \equiv g^2 E$, the
rescaled initial oscillator excitation number, $\nu \equiv g^2n $, 
(in the classical theory $n$ is {\it defined} as $E_{osc}/\omega$ and 
need not be integral)
and an initial oscillator phase, $\phi$.  The question at
this stage is whether the system can cross to the other side of the
barrier, {\it i.e.}~whether the transition is classically allowed.  In
particular, in the projection to the $\nu$-$\epsilon$ plane there will
be a classically 
allowed region where, for some value(s) of 
$\phi$, the system will evolve to the other side of the potential
barrier. The rest of the plane will consitute the classically
forbidden region where, no matter what the initial phase, the
system will bounce back from the barrier.

Clearly the entire domain $\epsilon < 1$ belongs to the classically
forbidden region: there can be no classical transition
with a total energy smaller than the barrier height (equal to $1$
in rescaled units).  However, a total energy larger than the barrier
height is {\it per se} no guarantee that the system will 
cross to the other side of the barrier.
The coupling between the center of mass and oscillator degrees
of freedom due to the potential will cause a transfer of
energy between the two, whose net effect can be repulsion from
the barrier even when the total energy is larger than the barrier height.
In general, for every initial value of $\nu$
there will be some minimal rescaled energy $\epsilon_0(\nu)$
such that for $\epsilon > \epsilon_0$ transitions across the 
barrier are possible.  The function $\epsilon_0(\nu)$ describes
the boundary of the classically allowed region.

The minimum of $\epsilon_0(\nu)$ is equal to 1 ({\it i.e.}~to the barrier
height).  Indeed, there is an obvious, unstable, static solution of
the equations of motion with both particles on top of the potential
barrier ($x_1(t)=x_2(t)=0$).  This solution, incidentally, corresponds to 
the static solution called the ``sphaleron'' in instanton mediated
processes \cite{Klinkhamer:1984di}.  If one perturbs this solution by giving an
arbitrarily small, common positive velocity to both particles, they
will move in the positive direction toward $X=\infty$.  (It is easy to
prove that the particles cannot go back over the barrier in this
situation.  If this were to happen, at some moment in time $x_1$ would
pass through zero.  At that moment, by conservation of energy, the
magnitude of the  center-of-mass velocity could not be larger than the
initial velocity of the particles.  But a perturbative analysis of the
initial motion shows that, however small its initial velocity may be,
the center of mass will acquire some finite positive momentum, which
will continue to increase so long as $x_1>0$. This implies that the
magnitude of the center of mass velocity cannot revert to its original
arbitrarily small value.)  Similarly, the time reversed
evolution has the two particles proceeding towards $X=-\infty$.  The
two evolutions, combined, describe therefore a classical process where
the system goes over the barrier with an energy larger, but
arbitrarily close to the barrier height.  This evolution, obtained by
an infinitesimal perturbation of the ``sphaleron'', will produce a
definite asymptotic value $\nu_0$ of the rescaled initial excitation
number, which will characterize the minimum of $\epsilon_0(\nu)$.

\begin{figure}[htbp]
  \begin{center}
  \includegraphics[clip,width=0.75\hsize]{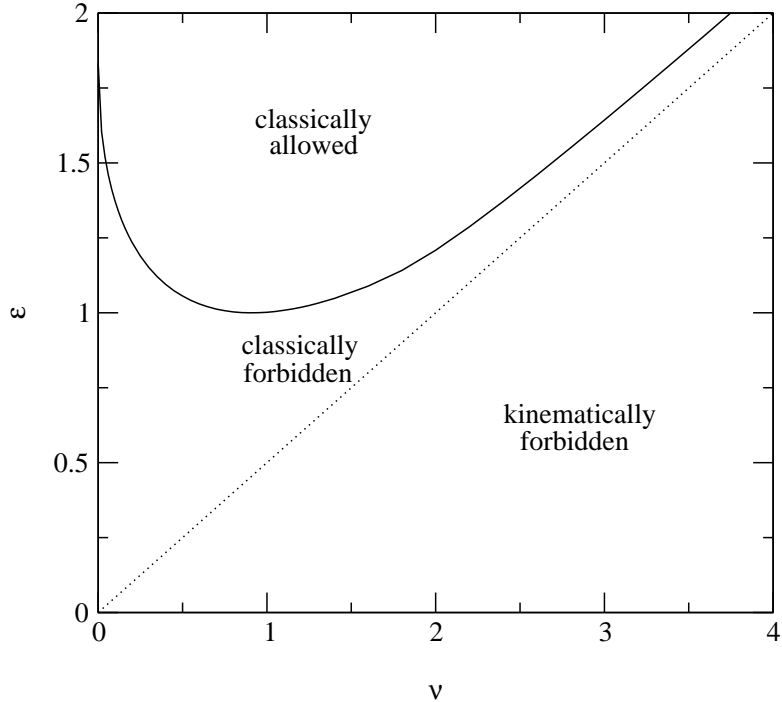}
  \caption{Boundary of the region of classically allowed transitions.}
  \label{fig:e-nu}
  \end{center}
\end{figure}

Values of $\nu$ smaller or larger than $\nu_0$ will give rise to 
values $\epsilon_0(\nu)$ larger than the barrier height.
Of particular interest for us is $\epsilon_0(0)$, i.e.~the lowest energy
for which one can have a classically allowed transition with the
incoming system in its ground state.  In the limiting cases
of infinite oscillator strength ($\omega \to \infty$) or
of decoupled particles ($\omega = 0$) $\epsilon_0(0)$ takes
values 1 and 2, respectively.  Indeed, the former case is equivalent to 
having a single particle (with twice the mass), which will evolve
over the barrier as soon as its initial energy is larger than the
barrier height.  In the second case, the particles proceed independently,
sharing the initial energy, and the center of mass will move to positive
infinity whenever the particle that feels the potential will be able to 
move over the barrier, which of course will happen when its share
of the total energy is larger than 1.  For finite, non-vanishing
values of $\omega$, the value of $\epsilon_0(0)$, or, more generally,
$\epsilon_0(\nu)$, must be determined numerically.
The small number of degrees of freedom in our model allows us to 
explore the phase space systematically, varying $\phi$, 
$\epsilon$ and $\nu$.  In this way we can determine
the boundary of the classically allowed region with reasonable accuracy.
Throughout this paper we will take $\omega=1/2$.  The 
corresponding boundary of the classically allowed region is illustrated
in Fig.~\ref{fig:e-nu}.  (The dotted line in the figure represents the
kinematic boundary $\omega\nu \le \epsilon$.)
With $\omega=1/2$, $\epsilon_0(0)$ equals approximately 1.8.
For energies lower than this value,  an incoming system in its
ground state may transit the barrier only through
tunneling.  In the next section we will present a full 
quantum-mechanical calculation of the corresponding transmission 
probability ${\cal T}_0(E,g)$.  In Sect.~\ref{sec:semi} we will adapt 
the semiclassical
formalism of Ref.~\cite{Rubakov:1992fb,Rubakov:1992ec} to 
derive an exponential expression
for the leading factor in ${\cal T}_0$, and calculate the exponent.

It is worthwile to mention that the existence of a definite
$\epsilon_0(0)$ where the boundary of classically allowed region
meets the axis $\nu=0$ is one important property which 
our simplified model, most likely, does not share with its more
complex field theoretical counterparts.  The infinite number of degrees
of freedom of the field theoretic systems opens the possibility
that the lower boundary of the classically allowed region approaches
the $\nu=0$ axis only asymptotically, or even that it is bounded below
by some non-vanishing minimum $\nu$ (cfr.~Ref.~\cite{Rebbi:1996zx} for a
numerical study of classically allowed transitions in the $SU(2)$-Higgs
system).

\section{Quantum-Mechanical Solution}
  \label{sec:quantum}

For the study of the full quantum system it is convenient to use the
basis formed by the tensor product of the center-of-mass coordinate
basis and the oscillator excitation number basis: $\vert X \rangle
\otimes \vert n \rangle$.  In this basis the state of the system
is represented by the multi-component wavefunction
\begin{equation}
  \label{eq:wfunct}
    \psi_n(X) \equiv (\langle X \vert \otimes \langle n\vert) 
    \vert \Psi \rangle
\end{equation}
and the time independent Schr\"odinger equation reads
\begin{equation}
  \label{eq:schroed}
    -{\partial ^2 \psi_n(X) \over \partial X^2} + \Big(n+{1 \over 2}\Big)
\omega \psi_n(X) + \sum_{n^\prime} V_{n,n^\prime}(X) \psi_{n^\prime}(X) = E \psi_n(X)
\end{equation}
where
\begin{equation}
  \label{eq:vnn}
    V_{n,n^\prime}(X)= \langle n\vert {1 \over g^2} e^{-{1 \over 2} g^2 (X+y)}
\vert n^\prime \rangle
\end{equation}
In the asymptotic region (large $|X|$) the interaction
terms are negligible, and the solution takes the form
\begin{equation}
  \label{eq:asym}
   \lim_{X\to\pm\infty} \psi_{n}(X)=t^{\pm}_n e^{\imath k_n X}+r^{\pm}_n e^{-\imath k_n X}
\end{equation}
with
\begin{equation}
  \label{eq:kn}
    k_n=\sqrt{E-\big(n+{1 \over 2}\big)\omega} 
\end{equation}
When  $n > E-\omega/2$, $k_n$ becomes imaginary; we
fix the continuation by defining
\begin{equation}
  \label{eq:knimag}
    k_n=\imath \sqrt{\big(n+{1 \over 2}\big)\omega-E}
\end{equation}

In order to calculate the transmission probability, we will look
for a solution with  
\begin{equation}
  \label{eq:cnneg}
    t^-_n=\delta_{n,0}
\end{equation}
and
\begin{equation}
  \label{eq:cnpos}
    r^+_n=0
\end{equation}
(This corresponds to an incoming system in its ground state.  The 
generalization to a process with an incoming excited state is
straightforward.) 
The inhomogeneous boundary conditions~(\ref{eq:cnneg}), (\ref{eq:cnpos}) 
fix 
the solution completely and the transmission probability is then given by
\begin{equation}
  \label{eq:transcoeff}
    {\cal T}_0= \sum_{n \le E/\omega - 1/2} {k_n \over k_0} |t^+_n|^2
\end{equation}
Numerical methods may be used to calculate the solution satisfying the
boundary conditions~(\ref{eq:cnpos}), (\ref{eq:cnneg}), 
and therefore
also ${\cal T}_0$.  In the rest of this section we
outline our computational procedure and describe the result. 

   To solve the Schr\"odinger equation numerically we must
discretize and truncate the system to leave
a finite, albeit very large, number of unknowns.
We accomplish this by replacing the continuum  $X$-axis
with a discrete and finite set of equally spaced vertices
\begin{equation}
  \label{eq:discrx}
    X_i=i a
\end{equation}
where $a$ denotes the lattice spacing and $-N_x \le i \le N_x$.
The truncation in oscillator space is perfomed by restricting
$n \le N_o$.

   To keep our notation concise, we will omit
the oscillator indices, using implicit vector and matrix notation,
and will use subscripts for the locations along the $X$ axis.
Thus, for instance, the expression $\sum_{n^\prime} V_{n,n^\prime}(ia) \psi_{n^\prime}(ia)$
will be simply written as $V_i \psi_i$.
It is also convenient to rewrite the continuum equation in the form
\begin{equation}
  \label{eq:schroedtwo}
    {\partial^2 \psi(X) \over \partial X^2} = A(X) \psi(X)
\end{equation}
where the matrix $A(X)$ is given by
\begin{equation}
  \label{eq:matrixa}
    A_{n,n^\prime}(X) = \Big[ \big(n+{1 \over 2}\big) \omega -E \Big] \delta_{n,n^\prime}
    +V_{n,n^\prime}(X)
\end{equation}

The discretization of Eq.~(\ref{eq:schroedtwo}) could be accomplished in a
straightforward manner by using the central difference approximation
of the second derivative with respect to $X$
\begin{equation}
  \label{eq:cda}
     \partial^2 \psi(X) /\partial X^2 = \frac{\psi(X+a)+\psi(X-a)
     -2\psi(X)}{a^2} + O(a^2)
\end{equation}
This would give the equations
\begin{equation}
  \label{eq:naive}
    \psi_{i+1}-2\psi_i+\psi_{i-1}=A_i \psi_i 
\end{equation}
with an error $O(a^4)$.

We have actually used the more sophisticated Numerov-Cowling
algorithm, which allows us to improve the accuracy of the discretization
by two powers of $a$.   From the Taylor series expansion 
of $\psi(X)$ we immediately find
\begin{equation}
  \label{eq:taylor}
    \psi_{i+1}+\psi_{i-1}-2\psi_i = 
    \Bigg[{\partial^2 \psi \over \partial X^2}\Bigg]_i a^2
    +\Bigg[{\partial^4 \psi \over \partial X^4}\Bigg]_i {a^4 \over 12}
    +O(a^6)
\end{equation}
Using Eq.~(\ref{eq:schroedtwo}), $(\partial^4  \psi/
\partial X^4) (a^4/12)$ 
can be written as $(\partial^2 A \psi/
\partial X^2) (a^4/12)$; this can be in turn approximated by
$(A_{i+1}\psi_{i+1}+A_{i-1}\psi_{i-1}-2 A_i\psi_i)(a^2/12)$ with the
same level of accuracy.  We are thus finally led to the following
discretization of eq.(\ref{eq:schroedtwo}):
\begin{equation}
  \label{eq:discsch}
    \psi_{i+1}-2\psi_i+\psi_{i-1}=
    {a^2 \over 12} A_{i+1} \psi_{i+1}  + {5 a^2 \over 6} A_i \psi_i 
    + {a^2 \over 12} A_{i-1} \psi_{i-1}
\end{equation}
which entails an error of order $a^6$.

Before proceeding further, we turn for a moment to the calculation
of the matrix elements  $V_{n,n^\prime}(X)$, which are needed for solving the
Schr\"odinger equation.  These can be calculated very efficiently
by means of a recursion procedure.  $V_{n,n^\prime}(X)$ is given by
\begin{equation}
  \label{eq:vnntwo}
    V_{n,n^\prime}(X)={1 \over g^2}\, \langle v_n\vert v_{n^\prime} \rangle 
\end{equation}
where $\vert v_n \rangle$ denotes the state
\begin{equation}
  \label{eq:vn}
    \vert v_n \rangle=  e^{-{1 \over 4} g^2 (X+y)^2}
    \vert n \rangle
\end{equation}
It is convenient to use the $y$ coordinate representation, writing
\begin{equation}
  \label{eq:vntwo}
    \vert v_n \rangle=  e^{-{1 \over 4} g^2 (X+y)^2}
    {(a^{\dag})^n \over \sqrt{n!}} \Big({\omega \over \pi}\Big)^{1/4}
    e^{-{1 \over 2} \omega y^2}
\end{equation}
with
\begin{eqnarray}
  \label{eq:aadagger}
    a={1 \over \sqrt{2 \omega}} {d \over dy} + 
    \sqrt{{\omega \over 2}} y \nonumber \\
    a^{\dag}=-{1\over\sqrt{2 \omega}}{d \over dy}+\sqrt{{\omega \over 2}}y
\end{eqnarray}
The first exponential in Eq.~(\ref{eq:vntwo}) can be brought to the right,
using
\begin{equation}
  \label{eq:movexp}
  e^{-{1 \over 4} g^2 (X+y)^2} a^{\dag}=
  \Big( a^{\dag} - {g^2 \over 2 \sqrt{2 \omega}} (X+y)\Big) 
  e^{-{1 \over 4} g^2 (X+y)^2} 
\end{equation}
This gives
\begin{equation}
  \label{eq:newvn}
    \vert v_n \rangle= {1 \over \sqrt{n!}} \Big({\omega \over \pi}\Big)^{1/4}
    \Big( a^{\dag} - {g^2 \over 2 \sqrt{2 \omega}} (X+y)\Big)^n 
    e^{-{1 \over 4} g^2 (X+y)^2 -{1 \over 2} \omega y^2}
\end{equation}
The exponential on the r.h.s.~may be written as a constant times
the ground state wavefunction of a shifted oscillator with 
frequency $\Omega$:
\begin{equation}
  \label{eq:newosc}
    e^{-{1 \over 4} g^2 (X+y)^2 -{1 \over 2} \omega y^2}
    =\Big({\pi \over \Omega}\Big)^{1/4}e^{- {g^2 \omega \over 4 \Omega}X^2} 
    \Big({\Omega \over \pi}\Big)^{1/4} e^{-{1 \over 2} \Omega z^2}
\end{equation}
with
\begin{equation}
  \label{eq:newfreq}
    \Omega=\omega+{g^2 \over 2}
\end{equation}
\begin{equation}
  \label{eq:newcoord}
    z=y+{g^2 \over 2\Omega} X 
\end{equation}
Re-expressing everything in terms of the creation and annihilation
operators for this new oscillator
\begin{eqnarray}
  \label{eq:bbdagger}
    b={1 \over \sqrt{2 \Omega}} {d \over dz} + 
    \sqrt{{\Omega \over 2}} z \nonumber \\
    b^{\dag}=-{1\over\sqrt{2 \Omega}}{d \over dz}+\sqrt{{\Omega \over 2}}z
\end{eqnarray}
we find
\begin{equation}
  \label{eq:newvntwo}
    \vert v_n \rangle= {1 \over \sqrt{n!}} 
    \Big({\omega \over \Omega}\Big)^{1/4}e^{- {g^2 \omega \over 4 \Omega}X^2} 
    \Big(\alpha b^{\dag} + \beta b + \gamma \Big)^n \vert 0 \rangle_b   
\end{equation}
with
\begin{eqnarray}
  \label{eq:coeff}
    \alpha&=&\sqrt{{\omega \over \Omega}} \nonumber \\
    \beta&=&-{g^2 \over 2 \sqrt{\omega \Omega}}  \nonumber \\
    \gamma&=&-{g^2 \over \Omega}\sqrt{{\omega \over 2}}\, X \ .
\end{eqnarray}
It is now straightforward to calculate the components of of $\vert v_n 
\rangle$ in the $b$-oscillator basis, and therefore also the inner products 
$\langle v_n \vert v_{n^\prime} \rangle$, by numerical iteration. 

In order to solve the discretized Schr\"odinger equation, we must
calculate $\psi_{i,n}$ (with the oscillator index explicit) 
for $-N_x \le i \le N_x$, $0 \le n \le N_o$.  This amounts to $2(N_x+1) N_o$ 
complex variables.  These satisfy Eq.~(\ref{eq:discsch}), 
for $\-N_x+1
\le i \le N_x-1$, for a total of $2(N_x-1) N_o$ complex conditions.
In addition, the boundary Eq.~(\ref{eq:cnneg}), (\ref{eq:cnpos})
give us the $2N_o$ complex conditions\footnote{In these equations
we can use either the $k_n$ given by the continuum dispersion formula
Eq.~(\ref{eq:kn}) or those given by the dispersion formula that 
follows from Eq.~(\ref{eq:schroedtwo}).  Given the high accuracy of the 
Numerov-Cowling discretization, the two options produce practically 
indistinguishable results.}
\begin{eqnarray}
  \label{eq:boundary}
    \psi_{-N_x,0}&=&e^{\imath k_0 a} \psi_{-N_x+1,0}+\big(1-e^{\imath k_0 a}
    \big)    \nonumber \\
    \psi_{-N_x,n}&=&e^{\imath k_n a} \psi_{-N_x+1,n} \quad\quad n>0 
    \nonumber \\
    \psi_{N_x,n}&=&e^{\imath k_n a} \psi_{N_x-1,n} 
\end{eqnarray}
Altogether, we thus have a system of $2(N_x+1) N_o$ complex, non-homogeneous
linear equations, which is precisely the number needed to determine all of
the unknowns. In order to insure good accuracy 
of the solution, we found it prudent to use 
cut-off values as large as $N_x=4096$ and $N_o=400$.
With such numbers  it would be impossible to tackle the system by 
brute force using a general purpose solver:  
this corresponds to a system of over 3 million complex equations,
which could not possibly be solved by a general purpose procedure.
However, we can take advantage of the special form of 
Eq.~(\ref{eq:discsch})  to implement an efficient 
solution procedure.  Indeed, by inverting
a set of $(N_o+1) \times (N_o+1)$ matrices, which is computationally
feasible, Eq.~(\ref{eq:schroedtwo}) can be recast in the form
\begin{equation}
  \label{eq:system}
    \psi_i=L_i \psi_{i-1} + R_i \psi_{i+1}
\end{equation}
where $L_i$ and $R_i$ are again $(N_o+1) \times (N_o+1)$ matrices.
The elimination of any definite $\psi_i$ now leads to a system
of equations for the remaining variables which, with $(N_o+1) \times (N_o+1)$
matrix algebra and matrix inversion, can be brought to the same form
of Eq.~(\ref{eq:system}), with suitably redefined $L$ and $R$ matrices.
We have used this procedure to progressively eliminate all the
intermediate variables $\psi_i$, $i=-N_x+1 \dots N_x-1$, ultimately
leaving a system of equations linearly relating all $\psi_i$ to 
$\psi_{-N_x}$ and $\psi_{N_x}$.  (Loosely speaking, the procedure can
be considered the implementation of a Green's function for our discretized
system of equations.)  In particular, $\psi_{-N_x+1}$ and $\psi_{N_x-1}$
are thus given as linear combinations of $\psi_{-N_x}$ and $\psi_{N_x}$.
Substituting these linear combinations in Eq.~(\ref{eq:boundary}) 
we now obtain
a system of $2 N_o +2$ complex, linear, non-homogeneous equations for the
$2 N_o +2$ complex variables $\psi_{-N_x}$, $\psi_{N_x}$, which can be
easily solved numerically.  As a final remark, we observe that 
the solution procedure outlined above only requires manipulation of
real matrices for the elimination of the intermediate variables, which
entails a substantial saving of memory and processor time.  Moreover,
we can also take advantage of the obvious symmetry under reflection
of the $X$-axis to further halve the computational costs.
\begin{table}[h]
{\footnotesize
\hskip -8.5mm
\begin{tabular} {|l| l l l l l l |r|} \hline
$g^2E$ & $g^2=0.01$ &$g^2=0.02$ &$g^2=0.03$ &$g^2=0.04$ &$g^2=0.06$ &$g^2=0.09$
& $-F$ \\ \hline
1.00 &              &               &             &               &
              & 0.00000241 &   -1.1520  \\
1.04 &              &               &             &               & 0.00000004
   & 0.00001080 &   -0.9906  \\
1.08 &              &               &             &               & 0.00000039
   & 0.00004266 &   -0.8446  \\
1.12 &              &               &             &  0.00000001   & 0.00000269
   & 0.00014584 &   -0.7143  \\
1.16 &              &               &             &  0.00000010   & 0.00001472
   & 0.00043282 &   -0.5477  \\
1.20 &              &               & 0.00000001  &  0.00000094   & 0.00006497
   & 0.00112833 &   -0.5051  \\
1.24 &              &               & 0.00000020  &  0.00000689   & 0.00023717
   & 0.00262162 &   -0.4254  \\
1.28 &              &               & 0.00000203  &  0.00003837   & 0.00073434
   & 0.00550586 &   -0.3603  \\
1.32 &              &  0.00000011   & 0.00001539  &  0.00017019   & 0.00197114
   & 0.01058551 &   -0.2948  \\
1.36 &              &  0.00000165   & 0.00008629  &  0.00062109   & 0.00467386
   & 0.01883914 &   -0.2375  \\
1.40 &              &  0.00001644   & 0.00038313  &  0.00191059   & 0.00994856
   & 0.03133460 &   -0.1889  \\
1.44 &  0.00000003  &  0.00011153   & 0.00139099  &  0.00506866   & 0.01926910
   & 0.04910559 &   -0.1625  \\
1.48 &  0.00000130  &  0.00057657   & 0.00423391  &  0.01182169   & 0.03435230
   & 0.07301430 &   -0.1220  \\
1.52 &  0.00002648  &  0.00234899   & 0.01103532  &  0.02460109   & 0.05692140
   & 0.10362365 &   -0.0897  \\
1.56 &  0.00026569  &  0.00775891   & 0.02508580  &  0.04629424   & 0.08840244
   & 0.14111159 &   -0.0675  \\
1.60 &  0.00181804  &  0.02128779   & 0.05054193  &  0.07972148   & 0.12963479
   & 0.18520851 &   -0.0492  \\
1.64 &  0.00886218  &  0.04956287   & 0.09149895  &  0.12687764   & 0.18064354
   & 0.23519177 &   -0.0344  \\
1.68 &  0.03199884  &  0.09978377   & 0.15067840  &  0.18832587   & 0.24056017
   & 0.28995637 &   -0.0227  \\
1.72 &  0.08854723  &  0.17673890   & 0.22824543  &  0.26278176   & 0.30770372
   & 0.34813584 &   -0.0138  \\
1.76 &  0.19415580  &  0.27975333   & 0.32121430  &  0.34722470   & 0.37974509
   & 0.40829435 &   -0.0073  \\
1.80 &  0.34839619  &  0.40164508   & 0.42390234  &  0.43740276   & 0.45411756
   & 0.46898820 &   -0.0028  \\
1.84 &  0.52777191  &  0.53035499   & 0.52913755  &  0.52842410   & 0.52816695
   & 0.52887867 &   -0.0001  \\ \hline
\end{tabular}
}
\caption{Results for the transmission probability.}
\end{table}
\begin{figure}[htbp]
  \begin{center}
    \includegraphics[clip,width=0.75\hsize]{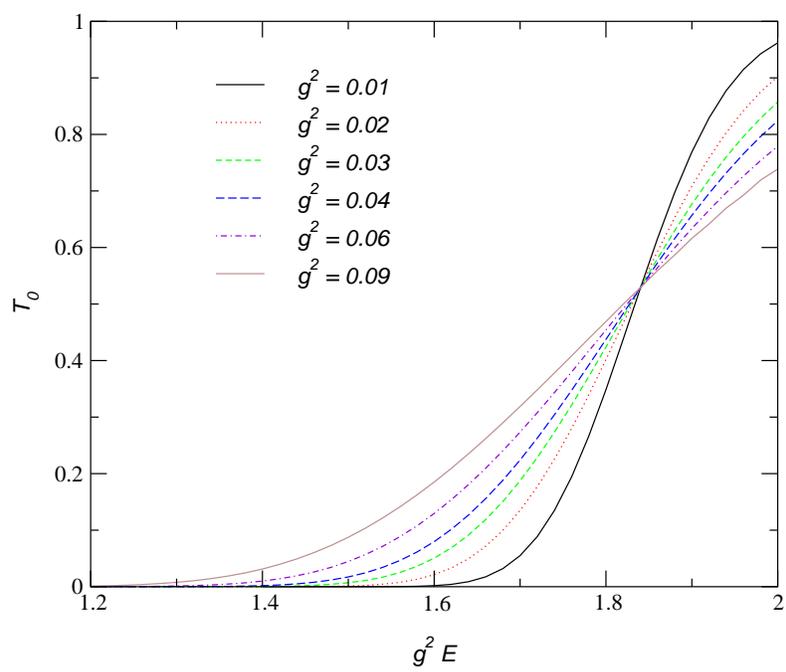}
    \caption{Transmission probability as function of rescaled total
      energy.}
  \label{fig:qm}
  \end{center}
\end{figure}

For our numerical calculations we have used $\omega=0.5$ for the
oscillator constant.  We have found this value a good middle ground
betweeen the extremes of very tight and very loose oscillator
coupling, where the novel features introduced by the internal degree
of freedom become less evident.  Also, apart from some calculations
where we varied parameters to study the effects of the discretization,
we have used a cut-off $N_x=2048$ and a lattice spacing $a=0.03
\sqrt{2}$.  
Insofar as $N_o$ is
concerned, we insured that its value is large enough that the
cut-off energy $(N_o+1/2) \omega$ exceeds the barrier height by
at least a factor of two. We have also checked
that the highest modes are essentially uncoupled.  Specifically, we
have used $N_o=400$ for $g^2=0.01$ and $g^2=0.02$ and $N_o=200$ for
all other values of $g^2$ (namely $g^2=0.03,\; 0.04,\; 0.06$ and
$0.09$).  We have solved the Schr\"odinger equation for values of $g^2
E$ ranging between $1$ and $2$ in steps of $0.02$.  A slightly thinned
out compilation of our data is presented in Table~1.

\begin{figure}[htbp]
  \begin{center}
    \includegraphics[clip,width=0.75\hsize]{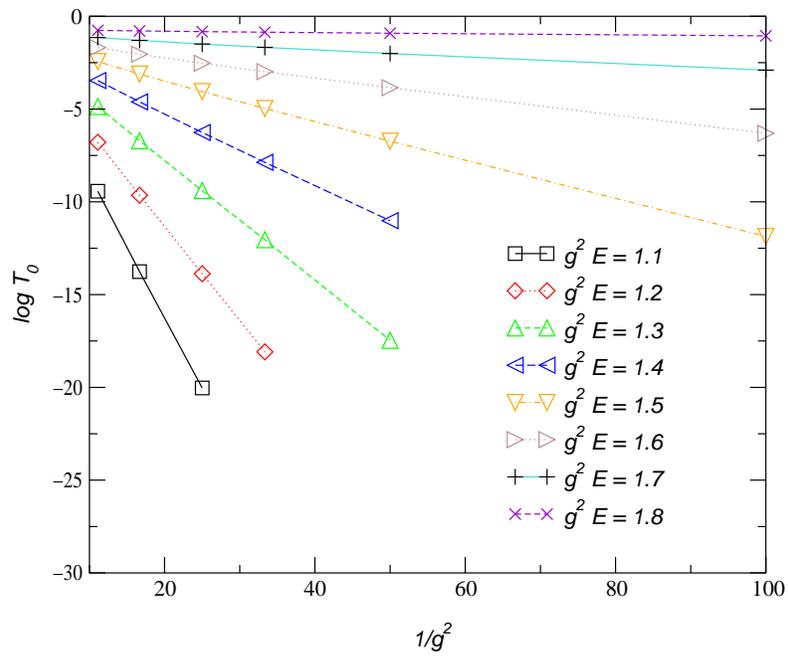}
    \caption{Logarithm of the transmission probability as function of
      $1/g^2$.}
  \label{fig:qmt}
  \end{center}
\end{figure}

Our results are also illustrated in Fig.~\ref{fig:qm}.  The approach to the
classical limit (a step function at $g^2 E=\epsilon_0(0)$) is evident.
In Fig.~\ref{fig:qmt} we plot the logarithm of the transmission probability as
function of $1/g^2$ for 8 values of $g^2 E$ equally spaced between 1.1
and 1.8.  The conjecture following from the semiclassical treatment is
that for small $g^2$ the logarithm of the transmission probability
should exhibit the linear behavior in $1/g^2$ at fixed
$g^2 E$  (cfr.~Eq.~(\ref{eq:transzero}))
\begin{equation}
  \label{eq:logt}
    \log {\cal T}_0=\log C_0(g^2 E) - {1 \over g^2} F_0(g^2 E)
\end{equation} 
This is well supported by the data in Fig.~\ref{fig:qmt}.  We have used the
slope of the last segment ({\it i.e.}~the segment corresponding to the
two largest 
$1/g^2$) for which we have significant data to derive the values of $F_0$
reproduced in Table~1.  In the next section we will compare them to 
the results of a semiclassical calculation. 

Finally, we performed several checks  to estimate the
accuracy of our numerical calculations.  For all solutions we verified
the degree to which the unitarity constraint 
\begin{equation}
  \label{eq:unitarity}
    \sum_{n \le E/\omega - 1/2} {k_n \over k_0} \big(|t^+_n|^2 +|r^-_n|^2) = 1
\end{equation}
was satisfied.  We found this equation fulfilled with an error
ranging from $10^{-6}$ to $10^{-5}$.  One might object that some
of the data for ${\cal T}_0$ in Table~1 are much smaller than this error.
This does not necessarily invalidate them, 
as our use of a  discretized Green
function may capture the correct exponential decays with relative,
rather than absolute errors, of the above order of magnitude.  
The regularity in even the smallest entries in the Table~1
supports this argument.  In any event, even discarding all values
of ${\cal T}_0 \le 10^{-5}$ one would still be left with a rich sample of data
verifying Eq.~(\ref{eq:logt}).

We have checked the effects of the cut-offs and of the finiteness of 
the lattice spacing by repeating the calculation (for $g^2 =0.03,\; 
g^2 E=1.7$) with different values of  $N_x$, $N_o$ and $a$.
\begin{table}
\center
\begin{tabular}{|l| l| l| l|} \hline
 $\;N_o$ & $\ \ a$ & $\ N_x$ & $\ \ \ \ {\cal T}_0$ \\ \hline 
 200 & 0.05  & 1024 & 0.18435 \\
 200 & 0.03  & 2048 & 0.18727 \\
 250 & 0.03  & 2048 & 0.18728 \\
 200 & 0.025 & 2048 & 0.18750 \\
 200 & 0.025 & 4096 & 0.18750 \\
 250 & 0.025 & 2048 & 0.18750 \\
 250 & 0.02  & 2048 & 0.18759 \\ \hline
\end{tabular}
\caption{Check of discretization effects.}
\end{table}
The results are reproduced in Table~2 which indicates that, apart from 
the case of a substantial increase in $a$, the relative errors due to the 
discretization are of order $10^{-3}$.

An alternative approach to the calculation of ${\cal T}_0$ consists in solving
the time-dependent Schr\"odinger equation.  One can simulate then
the collision of a wave packet against the barrier and measure directly 
the trasmission probability.  It is of course crucial to implement
a solution scheme which preserves the unitarity of the evolution
(up to numerical round-off errors).  We did follow this approach
in some earlier calculations, using a split operator technique to 
achieve a unitary evolution.  We obtained results consistent 
with our later calculations based on the time independent Schr\"odinger 
equation. However solving the time dependent Schr\"odinger equation 
proved much more (CP) time consuming than solving the time independent
one and so we abandoned the former method in favor of the
technique described in this section.

\section{The Semi-Classical Formalism}
  \label{sec:semi}
We begin this Section with the derivation of the semi-classical
procedure for calculating the exponent $F(g^2E, g^2n)$
of the transmission probability from the $n$-th excited state at total 
energy $E$, Eq.(\ref{eq:transn}).
Consider an incoming state of the form
\begin{equation}
  \label{eq:incoming}
   \vert E,n \rangle_{\delta} = \int dP^\prime \Phi_{P,\delta}(P^\prime)
   \vert P^\prime,n>
\end{equation}
where 
\begin{equation}
  \label{eq:p}
   P=\sqrt{2(E-\omega n)} \ ,
\end{equation}
\begin{equation}
  \label{eq:pp}
   \vert P^\prime,n> = \frac{1}{\sqrt{2\pi}} \int dX e^{\imath P^\prime X} 
   \vert X \rangle \otimes \vert n \rangle 
\end{equation}
is a simultaneous eigenstate of the center of mass momentum and oscillator
number, and $\Phi_{P,\delta}(P^\prime)$ is a momentum space wavefunction
with the following properties:
\begin{itemize}
\item it is sharply peaked for $P^\prime \approx P$, with a width of
  order $\delta$; 
\item it corresponds to an $X$-space wave packet which has support
  only for $X \ll 0$, well outside of the range of the potential.
\end{itemize}

With these definitions, the transmission probability is given by
\begin{equation}
  \label{eq:transprob}
    {\cal T}_n(E)=\lim_{\delta \to 0} \lim_{t_f-t_i \to \infty}
    \int_0^{\infty} dX_f \int_{-\infty}^{\infty} dy_f \,
    \vert \langle X_f, y_f\vert e^{-\imath H (t_f-t_i)} 
    \vert E,n \rangle_{\delta} \vert^2 
\end{equation}
This motivates us to calculate the matrix element
\begin{equation}
  \label{eq:matel}
    A(X_f,y_f,P,n)=\langle X_f, y_f\vert e^{-\imath H (t_f-t_i)} 
    \vert P, n \rangle
\end{equation}
Position-eigenstate matrix  elements may be evaluated in terms of a
path integral involving the classical action:
\begin{equation}
  \label{eq:pi}
    \langle X_f, y_f\vert e^{-\imath H (t_f-t_i)} \vert X_i,
    y_i\rangle = C \int \![dX][dy]\, e^{\imath S}
\end{equation}
where $C$ is a normalization constant, and the integration is over
paths satisfying $X(t_i)=X_i,\; y(t_i)=y_i$ and $X(t_f)=X_f,\;
y(t_f)=y_f$.  The amplitude (\ref{eq:matel}) is the convolution of 
the path integral (\ref{eq:pi}) with the eigenfunctions of the 
center-of-mass momentum and oscillator excitation number, 
$e^{\imath PX}$ and $\langle y \vert n \rangle$, respectively. 
$\langle y \vert n \rangle$ is conveniently represented in terms of
an integral over coherent state variables $z$ and $\bar{z}$. In 
this way we obtain
\begin{eqnarray}
    A(X_f,y_f,P,n) = \frac{1}{\sqrt{2 \pi}}
  \int\! dX_i\, dy_i\, e^{\imath PX_i} \int\! {dz d{\bar z} \over 2\pi
       \imath }\, e^{-{\bar z} z} {{\bar z}^n\over \sqrt{n!}} \nonumber \\
    \label{eq:coherent}
    e^{-\frac{1}{2} z^2 
      - \frac{1}{2} {\omega} y_i^2 +
      \sqrt{2\omega} z y_i} 
         \langle X_f, y_f\vert e^{-\imath H (t_f-t_i)} \vert X_i,
    y_i\rangle
\end{eqnarray}

The main idea, adapted from the method of 
Ref.~\cite{Rubakov:1992fb,Rubakov:1992ec},
is to set $E=\epsilon /g^2$, $n=\nu /g^2$ and take the limit
$g\to 0$ while holding $\epsilon, \nu$ fixed. 
Indeed, by rescaling the integration variables, we 
are then able to recast the matrix element in the form:
\begin{equation}
  \label{eq:int}
    A = C\int dX_i\,dy_i \int\! {dz d{\bar z} \over 2\pi i}\,
    \int [dX][dy]\, e^{-\frac{1}{g^2}\,\Gamma}
\end{equation}
with
\begin{equation}
  \label{eq:action}
    \Gamma = -\imath S +\frac{1}{2} z^2 +\frac{1}{2} \omega y_i^2
    -\sqrt{2\omega} z y_i + {\bar z} z - \imath pX_i - \nu \ln {\bar z} +
    \frac{1}{2} \nu(\ln{\nu}-1)
\end{equation}
where $p= g P$, $S$ is the classical action, 
and Stirling's approximation has been used for the
factorial\footnote{In order to keep our notation simple, we have used
  the same symbols ({\it i.e.}~$X_i, y_i$) for all rescaled
  integration variables. Note that $X_f$ and $y_f$ must also be
  eventually integrated upon, cfr.~Eq.~(\ref{eq:transprob}), and are
  rescaled as well.}.

This form for the matrix element $A$ is now suitable for a
semiclassical analysis at small $g$: we find stationary points
of $\Gamma$, and evaluate the integral in a gaussian approximation
about such points. The stationarity conditions (obtained by varying
$X(t), y(t), X_i, {\bar z}, z$, and $y_i$) are:
\begin{equation}
  \label{eq:deltaS}
 {\delta S \over \delta X(t)} = {\delta S \over \delta
      y(t)}= 0\;,\;\; \; t\neq t_i, t_f
\end{equation}
\begin{equation}
   \label{eq:momentumbc}
     {dX\over dt}\Bigg|_{t=t_i} = p  
\end{equation}
\begin{eqnarray}
  \label{eq:eom}
    {\bar z} z &=& \nu \nonumber \\
    \Bigl(\omega y_i + \imath {dy\over
      dt}\Bigg|_{t=t_i}\Bigr) &=& \sqrt{2\omega}\, z \nonumber \\
    \Bigl(\omega y_i - \imath {dy\over
      dt}\Bigg|_{t=t_i}\Bigr) &=& \sqrt{2\omega}\,{\bar z} 
\end{eqnarray}

\begin{figure}[htbp]
  \begin{center}
    \includegraphics[clip,width=0.75\hsize]{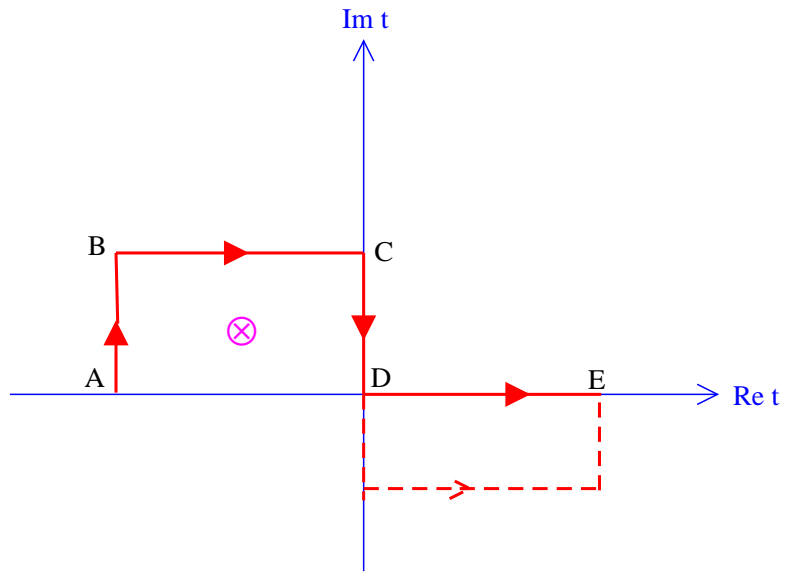}
    \caption{Contours in the complex time plane used to find the 
      saddle point solutions.}
  \label{fig:contour}
  \end{center}
\end{figure}

As expected, Eq.~(\ref{eq:deltaS}) is the classical equation of motion for
this system, while Eq.~(\ref{eq:momentumbc}) and (\ref{eq:eom})
imply that the initial classical state has the (rescaled)
center-of-mass momentum
$p$ and oscillator excitation number $\nu$, so that $\epsilon =
p^2/2 + \omega \nu$.
Therefore in the classically forbidden region
of the $\epsilon - \nu$ plane there will be no real solution
where the system goes over the barrier.
Nevertheless there may be {\em complex}
solutions. We expect that the integral is dominated by stationary
points, even if these points lie outside the domain of
integration. Hence we will seek solutions that may involve complex
values for the integration variables\footnote{In general this allows
values of $z$ and $\bar z$ such that ${\bar z} \ne z^*$.}. 
In searching for such solutions we must remember that we are performing
an analytic continuation of the integration variables; in general we
will run into singularities in the complex $t$-plane. To deal with this
problem we note that the {\em time} contour, the real time axis, can
be distorted into the complex plane without changing the path
integral, provided we keep the time contour end points ($t_i, t_f$)
fixed\footnote{The evolution operator may be written $\exp[-\imath H
    (t_f-t_i)] = \prod_j \exp[-\imath H {dt}_j]$ provided $\sum_j {dt}_j =
  (t_f - t_i)$. This argument applies even for complex
  ${dt}_j$.}. Thus our strategy will be to search for complex
solutions to Eq.~(\ref{eq:eom}) along a complex time contour 
$ABCDE$ as shown in
Fig.~\ref{fig:contour}.

The matrix element~(\ref{eq:matel}) in this approximation becomes 
\begin{equation}
  \label{eq:exp}
    A = e^{-\frac{1}{g^2} { \Gamma}(p,n,X_f, y_f) + c}
\end{equation}
where the correction $c$ determines  a pre-exponential factor in the
semiclassical limit, {\it i.e.} $\lim_{g\to 0} g^2 c = 0$. The function
$ \Gamma$ is equal to the right hand side of Eq.~(\ref{eq:action})
evaluated at  
the solution to
Eq.~(\ref{eq:deltaS}-\ref{eq:eom}). 
Using these equations to eliminate $z$ we write
\begin{equation}
  \label{eq:inv}
    {\Gamma} = -\imath S_0 - \imath \frac{1}{2} p X_i  -\nu \ln{\bar z} +
    \frac{1}{2} \nu\ln \nu \ 
\end{equation}
where
\begin{equation}
  \label{eq:s0}
    S_0= -\int_{t_i}^{t_f} \! dt\, \Biggl[ \frac{1}{2} X{d^2\over dt^2} X +
    \frac{1}{2} y({d^2\over dt^2} + \omega^2)y +
    e^{-(X+y)^2}\Biggr]+ \frac{1}{2} X_f {dX\over
    dt}\Bigg|_{t=t_f} + \frac{1}{2} y_f {dy\over
    dt}\Bigg|_{t=t_f} \ 
\end{equation}
Note that the quantity $S_0$ is insensitive to the value of
$t_i$, provided the solution to Eq.~(\ref{eq:deltaS}) 
is in the asymptotic
region near time $t_i$. 

Equations (\ref{eq:momentumbc}), 
(\ref{eq:eom}) and (\ref{eq:inv}) involve the quantities
defined at large negative time on the real time axis (region $A$
in Fig.~\ref{fig:contour}). It is convenient to formulate the boundary
conditions at large negative $\hbox{Re }t$ on the part $BC$ of the contour, 
where
\[
     t = t^\prime + \frac{\imath}{2}T\; , \;\;\;\; t^\prime = 
{\rm real} \to - \infty
\]
Since for the moment we consider the asymptotic past,
 we may ignore the
potential $V$ and write the solution at large negative $t^\prime$
as follows,
\begin{equation}
  \label{eq:bound5}
    y(t) = {1\over \sqrt{2 \omega}}(u e^{-\imath\omega t^\prime} + v
    e^{\imath\omega t^\prime}) 
\end{equation}
\begin{equation}
   \label{eq:bound6}
    X(t) = X_0 +   p t^\prime
\end{equation}
The three parameters of the solution, $X_0$, $u$ and $v$, which are in
general complex, are related to the quantities entering 
Eq.~(\ref{eq:momentumbc}),~(\ref{eq:eom}),~(\ref{eq:inv}) in an obvious way,
\[
    u e^{-\frac{1}{2}\omega T} = z e^{\imath \omega t_i}\;, \;\;\;
    v e^{\frac{1}{2}\omega T} = \bar{z} e^{- \imath \omega t_i}\;,
\]
\[
     X_i = X_0 - \frac{\imath}{2} pT + pt_i 
\]
The condition
\begin{equation}
     \label{eq:nuuv}
      \bar{z} z \equiv uv = \nu
\end{equation}
tells us that the phase of $u$ is opposite to
that of $v$, and we may parametrize them as
\begin{equation}
  \label{eq:theta5}
    v= e^{\theta} u^*
\end{equation}
So far we have not specified a value for the parameter $T$; we have
argued that our result is independent of this parameter, provided we
avoid singularities in the complex plane. Since variation of $T$
changes the value of $\hbox{Im } X_0$, we can adjust $T$ such that $X$
is real in the region $B$: $\hbox{Im } X_0 = 0$.

The transition probability is given by the absolute value of the
matrix element squared; that is, in terms of twice the real part of
$\Gamma$.
Using the above relations between $u$ and $v$ we can 
write the transmission probability as $\exp{(- F/g^2)}$ where
\begin{equation}
  \label{eq:prob}
    F = 2\hbox{Im } S_0  - \epsilon T - \nu \theta
\end{equation}
The resulting values of $T$ and $\theta$ depend on $\nu$ and
$\epsilon$. However, we may treat $T$ and $\theta$
as independent parameters instead, so that the boundary conditions
are formulated in a simple way in the asymptotic past on the part
$BC$ of the contour:

(i) $X(t^\prime)$ and $\dot{X} (t^\prime)$ are real at $B$.

(ii) positive and negative frequency parts of the oscillator 
solution (\ref{eq:bound5}) are related by Eq.~(\ref{eq:theta5})
at $B$.

At given $T$ and $\theta$, the initial center-of-mass momentum and 
excitation number (and hence the total energy)
are to be found from Eq.~(\ref{eq:momentumbc}) and
(\ref{eq:nuuv}).
It is straightforward to check that
\[
    \frac{\partial \left(2\hbox{Im } S_0 (T, \theta)\right)}{\partial T}
       = \epsilon
\]
\[
    \frac{\partial \left(2\hbox{Im } S_0 (T, \theta)\right)}{\partial \theta}
       = \nu
\]
so that $T$ and $\theta$ are Legendre conjugate to $\epsilon$
and $\nu$.
It is worth noting also
that  Eq.~(\ref{eq:theta5}) at $\theta \neq 0$
in fact {\it requires} the solution to be complex in the region B
of the contour.

We are interested in the total probability for transmission; thus we
should integrate our probability over all values of $y_f$ and
over positive values of $X_f$. This final integral may also be done
using the saddle point approximation. 
The saddle point condition is simply that 

 (iii) the solution $X(t)$ and $y(t)$ should be {\it real} along
the $D \to E$ part of the contour. 

At given $T$ and $\theta$, the 
classical equations of motion and the boundary conditions
(i), (ii) and (iii) are sufficient to specify the complex solution 
up to time translations along the real axis.  Finding the solutions
is still a non-trivial computational task.  To simplify this task we 
start from a sub-class of solutions with $\theta=0$, whose
numerical determination is easier,
and then deform these solutions to $\theta \ne 0$.

For the solutions with $\theta=0$
the $X$ and $y$ coordinates are analytic and real along 
the entire contour $BCDE$ of Fig.~\ref{fig:contour}.  From
the Cauchy-Riemann conditions it follows that the 
velocities $\dot X$ and $\dot y$ vanish at $C$ and $D$.  The
motion along the imaginary time axis can be reformulated
in terms of $\tau= \hbox{Im } t$ and  a ``Euclidean'' Lagrangian 
\begin{equation}
  \label{eq:euclideanlgn}
    L = \Biggl[\frac{1}{2} {\frac{d X}{d \tau}}^2 + \frac{1}{2} 
    {\dot{ y}}^2 + \frac{1}{2} \omega^2  y^2 + 
    e^{-\frac{1}{2}( X+ y)^2}\Biggr] 
\end{equation}
\begin{figure}[h!]
  \begin{center}
    \includegraphics[clip,width=0.75\hsize]{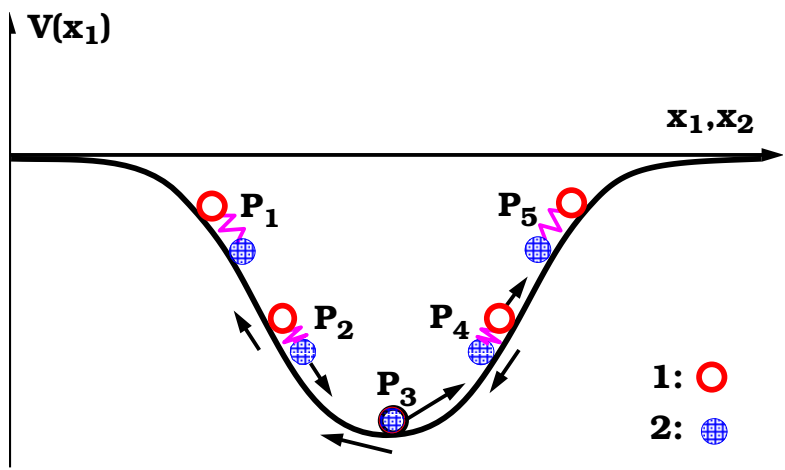}
    \caption{Motion in the ``periodic instanton'' solutions.}
    \label{fig:perinst}
  \end{center}
\end{figure}

The equations of motion
\begin{eqnarray}
  \label{eq:euclideaneqm}
    \frac{d^2 X}{d\tau^2} &=& -(X+y) e^{-\frac{1}{2}( X+ y)^2}\nonumber \\    
    \frac{d^2 y}{d \tau^2} &=& \omega^2 y -(X+y) e^{-\frac{1}{2}( X+ y)^2} 
\end{eqnarray}
describe the evolution of the system along the imaginary time
axis.  We look for periodic solutions where $dX/d\tau$
and $d y/d\tau$ vanish at the turning points
$\tau=0$ and $\tau=T/2$, $T$ being the
period of the motion.  These solutions are analogous to
the Euclidean solutions that are commonly used to describe
motion through a barrier in the semiclassical treatment of
tunneling with a single degree of freedom.
In this latter case finding
periodic solutions is straightforward: one need only integrate
the equations of motion with inverted potential.
The situation  with several degrees of
freedom is not so simple.  Indeed, the continuation to imaginary time
not only  
inverts the potential barrier, which now becomes a potential well,
but also changes the harmonic restoring force
into a linearly increasing repulsive force. This force makes the
system unstable, 
requiring careful adjustment of the values of $X$ and $y$ at the
turning points to obtain a periodic solution.
The resulting motion is similar to 
the ``periodic instanton'' solutions that appear in
topology changing transitions in quantum field theory \cite{Khlebnikov:1991th}.
There too, all of the field oscillator degrees of freedom
become repulsive in the Euclidean motion and the
solutions are unstable:
a small perturbation of the field profile at one of 
the turning points grows exponentially in the subsequent evolution.
The motion in the ``periodic instanton'' solutions of our model
is illustrated in Fig.~\ref{fig:perinst}.  At the turning
points ($P_1, P_5$) particle 1 is attracted towards the
bottom of the  potential well but repelled
by particle 2, which is located between particle 1 and the bottom
of the potential.  Both particles accelerate towards the
bottom (particle 2 because of the repulsive
force exerted by particle 1).  The balance of forces, however,
is such that particle 1 moves faster than particle 2, reducing
the interparticle distance ($P_2, P_4$) and, correspondingly, the repulsive
force, until it overtakes particle 2 precisely when
both particles transit through the bottom of the potential ($P_3$).
\begin{figure}[htbp]
  \begin{center}
  \includegraphics[clip,width=0.75\hsize]{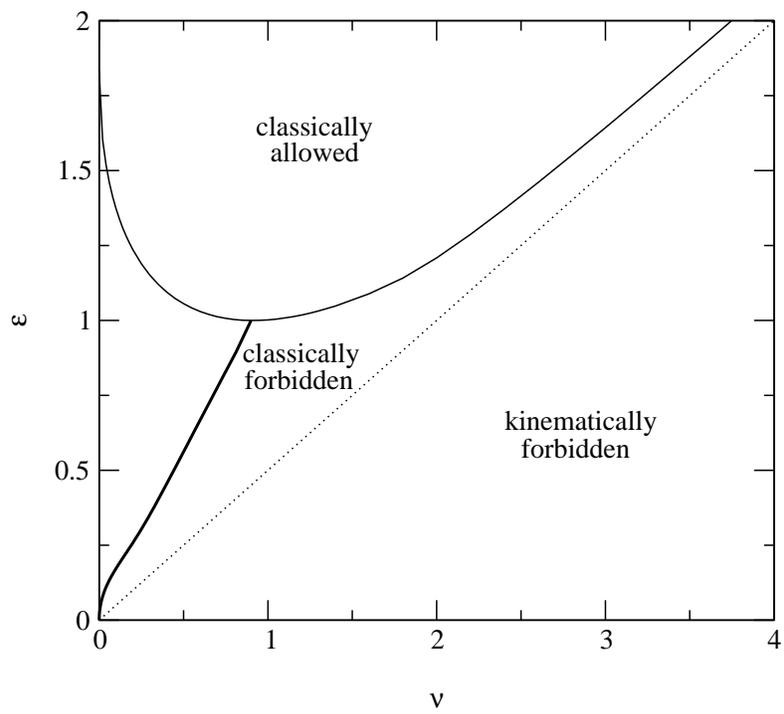}
  \caption{The curve spanned by the periodic instanton solutions.}
  \label{fig:perinstline}
  \end{center}
\end{figure}

Finding the periodic instanton solutions of
our model is rather easy.  We divide
the interval $0 \le \tau \le T/4$ into $N$ subintervals of
width $\Delta \tau = T/(4N)$ and we denote by $X_i$, $y_i$
the values taken by $X$ and $y$ at $\tau= i \Delta \tau$.
We discretize the Euclidean action $S_E=\int L_E \, d\tau$ 
and look for a minimum of $S_E$ with respect to the variables
$X_i, y_i, \;\; i=0 \dots N-1$, while $X_N$ and $y_N$ are
kept fixed at zero.  Since the Euclidean action
is bounded from below, the algorithm
of conjugate gradients converges rapidly to the correct solution.
The values $X_0, y_0$ can then be used
as initial data for the integration of the equations of
motion along the real time axis (from $D$ to $E$ in 
Fig.~\ref{fig:contour}) until both particles are far
out of range of the potential.  In this manner one can find
the asymptotic oscillator number of the solutions (in the
periodic instanton solutions initial and final oscillator numbers
are of course identical).  The periodic instanton solutions
span the one-dimensional subspace denoted by the thick line
in the $\nu$-$\epsilon$ plot of Fig.~\ref{fig:perinstline}.

Starting from the periodic instanton solutions we find
solutions with $\theta>0$ and reduced incoming oscillator number
$\nu$ by a deformation procedure.
In order to solve the equations of motion numerically we
subdivide the contour
$BCD$ of Fig.~\ref{fig:contour} into $N$ subintervals separated
by vertices labeled by an index $i=0 \dots N$.  We denote by
$\delta t_i$ the time interval from vertex $i$ to vertex $i+1$.
$\delta t_i$ will be real 
along the $BC$ portion of the contour,
after which $\delta t_i$ should be negative imaginary.  It is convenient,
however, to place the very last subinterval along the real time
axis.  Thus we place the vertex $N-1$ at the origin, and the vertex
$N$ at the point $t=\delta t_{N-1} = {\rm real}$.
The discretized action is
\begin{eqnarray}
    S & = & \sum_{i=0}^{N-1}\Biggl[\frac{(X_{i+1}-X_i)^2}{2\delta t_i}
    + \frac{(y_{i+1}-y_i)^2}{2\delta t_i} \nonumber \\
    & &-\omega^2 \frac{ y_{i+1}^2+y_i^2}{4}\, \delta t_i - 
    \frac{e^{-\frac{1}{2}(X_{i+1}+ y_{i+1})^2}
    +e^{-\frac{1}{2}(X_i+ y_i)^2}}{2}\,\delta t_i \Biggr] 
  \label{eq:disact}
\end{eqnarray}
This expression is quite general and valid for any
contour of integration in the complex time plane.

It is convenient to use $z_{i,j}$, $j=1,\dots, 4$, to denote the four
variables  
$\hbox{Re } X_i$, $\hbox{Im } X_i$, $\hbox{Re } y_i$, $\hbox{Im } y_i$. 
The equations of motion are given by
\begin{equation}
  \label{eq:motion}
    \frac{\partial S }{\partial z_{i,j}}=0 \quad \quad i=1 \dots N-1 
\end{equation}
These amount to $4N-4$ conditions for the $4N+4$ unknowns $z_{i,j}$.
The solution must also satisfy the boundary conditions
\begin{equation}
  \label{eq:bc1}
    y_0+y_1 -\frac{\imath}{\omega}(y_1-y_0)=
    e^{\theta}\Big[y_0^*+y_1^* -\frac{\imath}{\omega}(y_1^*-y_0^*)\Big]
\end{equation}
[cfr.~Eq.~(\ref{eq:bound5}), (\ref{eq:nuuv})] and
\begin{equation}
  \label{eq:bc2}
    \hbox{Im }X_{N-1}=\hbox{Im }X_N=\hbox{Im }y_{N-1}=\hbox{Im }y_N=0  \ .
\end{equation}
In addition, we remove the invariance under time translation
by demanding that $X_0$ takes a fixed real value
\begin{equation}
  \label{eq:bc3}
    X_0=c
\end{equation}
The precise value of $c$ is not relevant.  The only important criterion
that $c$ must satisfy is that the imaginary time axis falls between
the expected singular points of the solution.  The value 
of $c$ can be readjusted,
if necessary, so that the point in the complex time plane where ${\rm
  Re } X=0$  belongs to the $CD$ part of the contour.

Equations (\ref{eq:bc1})-(\ref{eq:bc3}) provide the required
8 additional conditions on the variables $z_{i,j}$.  We will
write these equations as 
\begin{equation}
  \label{eq:bc}
    B_k(z_{i,j})=0
\end{equation}

Starting from the periodic instanton solutions and evolving
them further from $C$ to $B$ we obtain an initial
class of solutions to Eq.~(\ref{eq:motion}), (\ref{eq:bc}) with $\theta=0$.
If we perform a small change of either $\theta$ or $T$, the two
parameters which indirectly determine $\nu$ and $\epsilon$, the
field configuration $z_{i,j}$ will no longer satisfy the equations
of motion.  We seek a correction $\delta z_{i,j}$ such that
$z_{i,j}+\delta z_{i,j}$ obey the equations of motion with
new values for $\theta$ and/or $T$:
\begin{equation}
  \label{eq:new1}
    \frac{\partial S }{\partial z_{i,j}}\Bigg\vert_{z+\delta z}=0
\end{equation}
\begin{equation}
  \label{eq:new2}
    B_k(z_{i,j}+\delta z_{i,j})=0
\end{equation}
If the deformation of the original solution is not too
large, Eq.~(\ref{eq:new1}), (\ref{eq:new2}) can be solved
by the Newton-Raphson method:  we expand to first order in
$\delta z$ and solve the linearized equations 
\begin{equation}
  \label{eq:linear1}
     \sum_{i',j'}\frac{\partial^2 S }
     {\partial z_{i,j}\partial z_{i',j'}}\Bigg\vert_z
      \delta z_{i',j'}= -\frac{\partial S }{\partial z_{i,j}}\Bigg\vert_z
\end{equation}
\begin{equation}
  \label{eq:linear2}
    \sum_{i',j'} \frac{\partial B_k}{\partial z_{i',j'}}\Bigg\vert_z
    \delta z_{i',j'}=-B_k\Big\vert_z
\end{equation}
This procedure is repeated until it converges to a solution.

In our calculations we typically took $N=2048$ and used the following
computational strategy.  Equation~(\ref{eq:linear1}) with a definite
index $i$ only couples the variables $\delta z_{i',j}$ with
$i'= i-1,\; i,\; i+1$.  It is then possible to use an elimination procedure
similar to the one outlined in Sect.~\ref{sec:quantum} (see
Eq.~(\ref{eq:system}) and considerations that follow) and express
all variables $\delta z_{i,j}$ in terms 
of $\delta z_{0,j}$, $\delta z_{N,j}$. (In practice this can be
done maintaining complex variables notation, which simplifies 
the arithmetic.  One must work with the explicit real and imaginary
parts of the variables only at the next stage of the calculation.)
Finally $\delta z_{0,j}$, $\delta z_{N,j}$, and
$\delta z_{1,j}$, $\delta z_{N-1,j}$ which, by virtue of the
elimination procedure are now expressed as linear functions
of $\delta z_{0,j}$, $\delta z_{N,j}$, are inserted into
Eq.~(\ref{eq:linear2}).  These equations thus become
a system of 8 real, linear, non-homogeneous equations in the
8 real variables $\delta z_{0,j}$, $\delta z_{N,j}$, that
can be straightforwardly solved. 
We are then able to start from $\theta=0$ (periodic instanton
solution) and gradually increase $\theta$ to a very large value,
which makes the incoming oscillator number $\nu$ effectively zero.
At the same time we gradually reduce the value of $T$, which
has the effect of increasing the energy $\epsilon$.  It is 
important to check that the solutions correspond indeed to tunneling 
processes, namely that in the further evolution along the positive real
time axis the center of mass coordinate $X$ goes to $+\infty$.
We found this to be the case up to $\epsilon \sim 1.1$.
At that point, though, the Newton-Raphson method develops
an instability and, when convergence is eventually reached,
further evolution along the real time axis shows a bounce from
the barrier with $X \to -\infty$.  We attribute this difficulty to 
the proximity of solutions with $X \to +\infty$ and $X \to -\infty$,
with possible bifurcation points.  In order to avoid falling
into a solution without tunneling, we continue a tunneling solution to 
a positive real value of $t$ along a contour extending
into $\hbox{Im }t < 0$, as illustrated by the dashed line in 
Fig.~\ref{fig:contour}.  At the final value of $t$ (point $E$ in the 
graph of Fig.~\ref{fig:contour}) the system is far into the positive $X$
domain and we can further increase $\epsilon$ without running
into any instability.

\begin{figure}[ht]
  \begin{center}
    \includegraphics[clip,width=0.75\hsize]{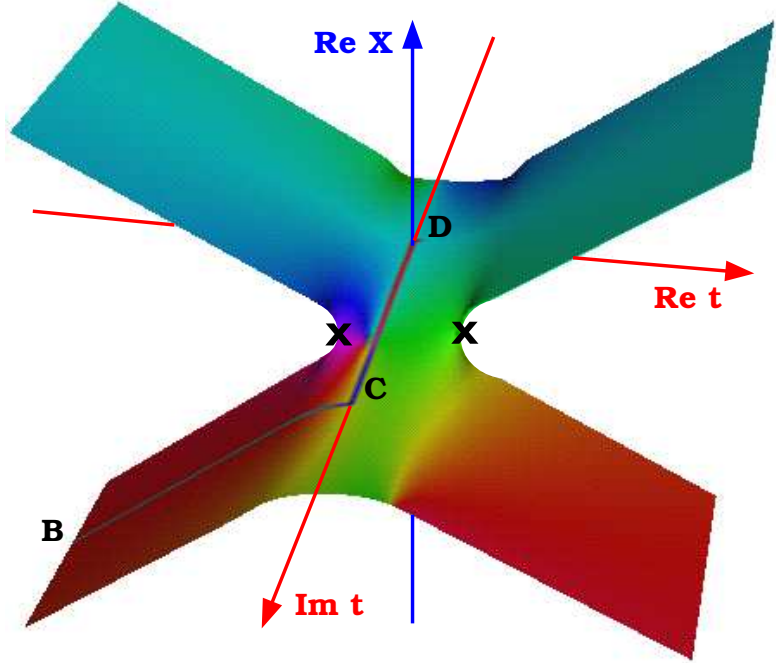}
    \caption{Tunneling solution to the equations of motion: center-of-mass 
      coordinate as function of complex time.  The singularities 
      (branch points) are labeled by crosses.}
  \label{fig:solution}
  \end{center}
\end{figure}

We illustrate in Fig.~\ref{fig:solution} a typical tunneling solution
in the complex time plane.  The figure displays the center of
mass coordinate $X$ as function of $\hbox{Re } t$, $\hbox{Im } t$
(we inverted the direction of the $\hbox{Im } t$ axis for a better
perspective).  The height of the surface gives the value of $\hbox{Re } X$,
while the phase of the complex variable $X$ is coded by color
(red for real negative, blue for real positive, with the other
values of the complex phase arranged in rainbow pattern---the color
will appear as gray-scale in a black and white printout). 
The contour of integration of the equations of motion is indicated
by a line of different color drawn on the surface.  The continuation
of $X$ to the entire complex plane has been obtained by starting
form the values along the imaginary time axis and integrating the
equations of motion outward with the leapfrog algorithm.
The singularities in the solution are quite apparent from 
Fig.~\ref{fig:solution}.  We found it noteworthy that one can
determine the singularity structure of the solutions 
numerically, since ordinarily one would expect numerical integration
methods to fail in the presence of a singularity.
The integration algorithm becomes unstable and diverges
as one approaches a singularity.  However it is possible to
exhibit the singularity structure by numerically integrating
the solution along closed contours around the singularities.  
Integration of the equations of motion by the leapfrog algorithm
along a closed contour entirely contained within a domain of
analyticity produces final values for $X$ and $y$ identical to 
the initial values to high degree of numerical accuracy,
equal to the expected discretization error $O((\delta t)^2)$
of the algorithm, whereas an enclosed singularity
is clearly present when the initial and final values of $X$ and $y$
are different.

\begin{table}
\center
\begin{tabular}{||l| l| l|| l| l| l||} \hline
$\;T/2$ & $\quad \epsilon$ & $\quad F$&$\;T/2$ & $\quad \epsilon$ & 
$\quad F$ \\ \hline 
1.75   & 1.0463  & 0.9715 & 0.35   & 1.4869  & 0.1038  \\
1.72   & 1.0846  & 0.8386 & 0.3    & 1.5208  & 0.0817  \\
1.7    & 1.1223  & 0.7103 & 0.25   & 1.5585  & 0.0611  \\
1.6    & 1.1334  & 0.6734 & 0.2    & 1.6005  & 0.0422  \\
1.4    & 1.1595  & 0.5950 & 0.175  & 1.6234  & 0.0336  \\
1.2    & 1.1921  & 0.5103 & 0.15   & 1.6477  & 0.0257  \\
1.0    & 1.2333  & 0.4195 & 0.125  & 1.6736  & 0.0186  \\
0.8    & 1.2867  & 0.3235 & 0.1    & 1.7011  & 0.0124  \\
0.7    & 1.3196  & 0.2741 & 0.075  & 1.7306  & 0.0073  \\
0.6    & 1.3578  & 0.2244 & 0.05   & 1.7621  & 0.0034  \\
0.5    & 1.4027  & 0.1749 & 0.0025 & 1.7960  & 0.0008  \\
0.4    & 1.4560  & 0.1268 & 0.01   & 1.8176  & 0.0001  \\ \hline
\end{tabular}
\caption{Results of the semiclassical analysis.}
\label{table:sc}
\end{table}

\bigskip

\begin{figure}[h!]
  \begin{center}
    \includegraphics[clip,width=0.75\hsize]{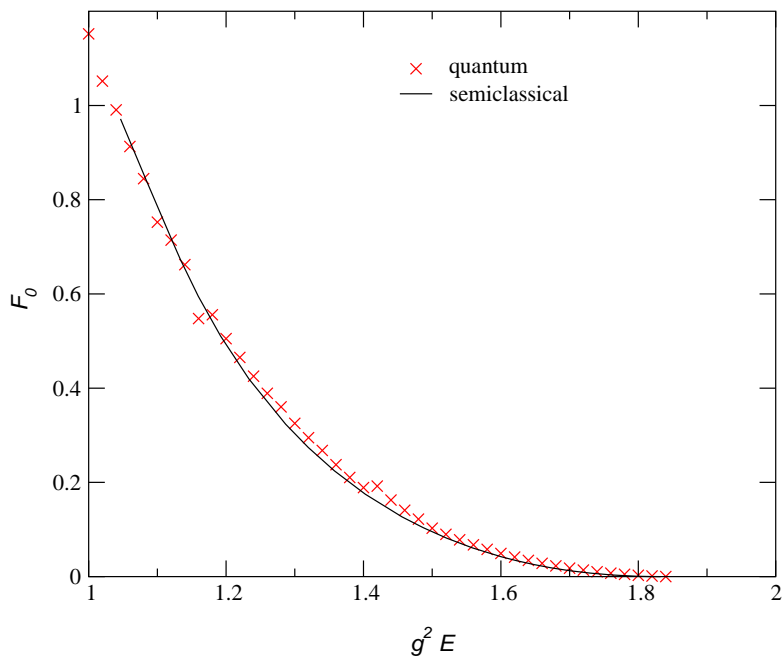}
    \caption{Comparison of the quantum mechanical and semiclassical results.}
  \label{fig:compare}
  \end{center}
\end{figure}

We reproduce in Table~\ref{table:sc} the results of our semiclassical
calculation.  The data correspond to $\theta=13$.
In Figure~\ref{fig:compare} we present a comparison of
the results for the exponent $F_0$ in the transmission probability 
(cfr.~Eq.(\ref{eq:twotoany})) obtained with the
full quantum-mechanical calculation (x) and with the semiclassical
technique (solid line).  In the quantum-mechanical calculation
we extracted $F_0$ from the slope of the last segment in the
graph of $\log {\cal T}_0$ versus $1/g^2$ at given energy for which we
had meaningful 
data (see Fig.~\ref{fig:qmt}). As a consequence, the line defined by the
crosses in Fig.~\ref{fig:compare} exhibits some small discontinuities.
We take the magnitude of these discontinuities as an indication
of the systematic errors in the quantum-mechanical calculation
due to the neglect of perturbative $O(g^2)$ and higher order effects.
Within these errors, the agreement between the results of the
full quantum-mechanical calculation and of the semiclassical calculation
is excellent.

\section{Conclusions}
  \label{sec:conclusion}
Our results validate, in the context of a model calculation,
the scaling formula of Eq.~(\ref{eq:twotoany}), (\ref{eq:transzero}),
the applicability of the method of Ref.~\cite{Rubakov:1992fb,Rubakov:1992ec} 
and the assumption that the ground
state transition probability can be obtained as the limit
of a more general transition probability from a coherent
initial state.

At the same time our investigation has brought to light
interesting properties of the analytic continuation of classical
solutions to complex time and complex phase space.
While the extension of classical motion to the complex time
domain has long formed the mainstay of semiclassical
calculations of tunneling, we believe that our specific
application shows novel features of the
analytically continued solutions intimately connected to 
the presence of several degrees of freedom.  Of particular
relevance we find that one can obtain information
on the singularity structure of the solutions by numerical
techniques.

With the qualification 
that a field has an infinite number of degrees of freedom
while our model has only two, our results bode well
for the application of the technique of 
Ref.~\cite{Rubakov:1992fb,Rubakov:1992ec} to field 
theoretical processes.
Hopefully, they will also open the path to new, imaginative
applications of semiclassical methods in other challenging
quantum-mechanical problems.

\bigskip
{\bf Ackowledgements.}  
The authors are indebted to
P.~Tinyakov for helpful discussions.
This research was supported
in part under DOE grant DE-FG02-91ER40676, RFBR grant
96-02-17449a, and by the
U.S.~Civilian Research and Development Foundation for
Independent States of FSU (CRDF) award RP1-187.
Two of the authors (C.R. and V.R.) would like to thank Professor
Miguel Virasoro 
for hopsitality at the Abdus Salam International Center for Theoretical
Physics, where part of this work was carried out.

\bibliographystyle{unsrt}

\end{document}